\documentstyle[12pt]{article}
\begin{document}
\title{\large\bf Charged Topological Black Hole Pair Creation}
\author{R.B. Mann$^a$  \\
Department of Physics \\
University of Waterloo, Waterloo, Ontario, Canada N2L 3G1\\ 
$^a${\it mann@avatar.uwaterloo.ca} \\ }
\date{\today \\ WATPHYS-TH97/05}
\maketitle
\begin{abstract}
I examine the pair creation of black holes in spacetimes with a cosmological
constant of either sign. I consider cosmological C-metrics and show that
the conical singularities in this metric vanish only for three distinct classes
of black hole metric, two of which have compact event horizons on each spatial
slice. One class is a generalization of the Reissner-Nordstrom
(anti) de Sitter black holes in which the event horizons are the direct
product of a null line with a 2-surface with topology of genus $g$. The other
class consists of neutral black holes whose event horizons are the direct 
product of a null conoid with a circle.  In the presence of a domain wall, 
black hole pairs of all possible types will be pair created for a wide range 
of mass and charge, including even negative mass black holes. I determine the 
relevant instantons and Euclidean actions for each case.
Only for spherical are non-static solutions possible.
\end{abstract}

\pagebreak
\section{Introduction}
\label{intro}

Pair creation of black holes continues to afford us interesting insights 
into quantum gravity and the relationship between entropy
and the number of quantum states of a black hole. It is
a tunnelling process in which the mass-energy of the 
created pair of black holes is balanced by their negative potential energy
in some  background field, such as that of an electromagnetic field 
\cite{dgkt}, a positive cosmological constant \cite{cosern},
a cosmic string \cite{strpair} or a domain wall \cite{dompair}. The
amplitude for the process is approximated by $e^{-I_i}$, where
$I_i$ is the action of the relevant instanton {\it i.e.} a Euclidean
solution to the field equations which interpolates between the states
before and after a pair of black holes is produced.

The $C$ metric solution of the Einstein-Maxwell equations
\cite{gen} is interpreted as describing two oppositely-charged black holes
undergoing uniform acceleration. It contains conical singularities, which
in general cannot be eliminated at both poles. These singularities are
interpreted as representing ``rods'' or ``strings'' which provide the
force necessary to accelerate the black holes.

This raises the question as to what can provide
the necessary force to accelerate the black holes. The simplest way
is to remove the conical singularities by
setting the acceleration to zero, in
which case the $C$ metric reduces to the Reissner-Nordstr\"om metric.
A less trivial means of removing the conical singularities is to allow the
black hole and acceleration horizons to coincide \cite{dgkt}; the 
Euclidean section of this metric has been referred to as the Type I instanton 
and has topology $S^2\times R^2$.
In this case the $C$ metric has the same functional form as
the Reissner-Nordstr\"om metric, with the coordinate that is usually
regarded as the azimuthal coordinate playing the role of
time. Analytically continuing $\phi \to i \Phi$ in the 
Euclidean Reissner-Nordstr\"om metric yields the type I instanton,
which can be interpreted as 
representing the decay of a $(2+1)$-dimensional
Kaluza-Klein vacuum,  $\mbox{Mink}^{(2,1)}\times S^1$ \cite{cosern}.

All other approaches have entailed adding additional forms of 
stress-energy to provide the necessary accelerating force. This typically
generalizes the $C$ metric to some other form, and it is the conicial
singularities of this modified metric that are then removed by some
limiting procedure on its parameters. The first such
approach taken \cite{ernst} added a
background electromagnetic field to the $C$ metric using a Harrison
transformation. This yielded the Ernst metric, which could be made regular
by a suitable choice of electromagnetic field strength, providing
the force necessary to accelerate the black holes \cite{dgkt,dggh}. 
It was subsequently
shown that a cosmological constant can also do the job \cite{cosern}, 
the conical singularities being eliminated by fixing the acceleration
parameter in terms of the cosmological constant. The resultant class
of regular metrics are all of the Reissner-Nordstr\"om de Sitter type.
Subsequent approaches have employed cosmic strings \cite{strpair} and
domain walls \cite{dompair} and dilatonic generalizations \cite{dggh,entar}
to provide the necessary accelerating force.

In a recent paper \cite{cqgtop} I demonstrated that the cosmological
$C$-metrics \cite{gen} contain a rich array of Euclidean instantons that
mediate pair production of black holes whose topology is of arbitrary
genus.  In this paper I consider a more detailed investigation of
how these black holes can be derived from the C-metric and of their
pair creation. The genus zero class of solutions has been studied before: 
it corresponds to the set of Reissner-Nordstr\"om de Sitter instantons that
mediate cosmological black hole pair production \cite{cosern}.
All of the higher genus solutions are asymptotically anti-de Sitter,
and correspond to instantons that are 
4 dimensional generalizations of the 3 dimensional black hole \cite{btz}.
Pair production of these black holes can take place in the presence of
domain walls whose topology is the same as that of the produced black
hole pairs. In a full theory of quantum gravity in which topology
changing processes are expected to be important, these 
instantons will have to be taken into account. 

The higher genus black hole spacetimes related to these instantons are 
of some interest in their own right.  Like their topologically spherical
cousins, they can form from gravitationally collapsing dust;
however unlike their cousins, the  initial density  of the collapsing
matter must be sufficiently small for collapse to take place \cite{wendy}.
A more recent investigation \cite{BLP} of their thermodynamic behaviour has indicated
that their entropy is one quarter their horizon area (although
this result appears to be sensitive to the subtraction procedure used to
make the Euclidean action finite \cite{vanzo}), and that they are thermodynamically
stable.  The ADM mass parameter of these black holes
can even be zero \cite{wendy,amin} or negative \cite{negmas}; the negative
mass hole can also form from the collapse of dust which violates the weak
energy condition \cite{negmas}. 

In section \ref{cosmoc} I consider the cosmological form of the C-metric, 
and show that removal of its conical singularities yields the full array 
of topological black hole spacetimes noted above. I also find that two 
other qualitatively different black hole spacetimes can emerge, both of 
which can be considered generalizations of the spacetimes associated with 
the type-I instanton mentioned above. One of these has a non-compact event 
horizon. The other is a generalization of the $(3+1)$ dimensional constant
curvature black hole recently considered by Banados \cite{Bancon}.
In section \ref{tads} I discuss the basic properties of these topological
black hole spacetimes, and compute their quasilocal mass and charge in
the large-radius (ADM) limit.  In section \ref{instsec} I consider
the pair creation of topological black holes using the domain wall
mechanism \cite{dompair,Chamsugra}. All topological black holes except
those of spherical topology can only be pair-created in a static configuration. 
The pair-production rate for a black hole of arbitrary genus 
is computed in section \ref{pair}. Constant curvature black hole pair
creation is considered in section \ref{conscurv}.  A 
few concluding remarks are contained in the final section.

\section{Cosmological $C$ metrics}
\label{cosmoc}

The cosmological $C$ metric can be written as
\begin{equation} \label{ccmet}
ds^2 = \frac{1}{A^2 (x-y)^2} \left[ H(y) dt^2 - H^{-1}(y) dy^2 +
G^{-1}(x) dx^2 + G(x) d\varphi^2 \right],
\end{equation}
where 
\begin{equation}\label{Gdef}
G(x) = \tilde{a}  - bx^2 -2mAx^3 -q^2A^2x^4
\end{equation}
and 
\begin{equation}\label{Hdef}
H(y) = a  - by^2 -2mAy^3 -q^2A^2y^4
\end{equation}
with $\tilde{a} = a - \frac{\Lambda}{3A^2}$, $\Lambda$ being the
cosmological constant and $A$ the acceleration parameter.
The gauge field in the magnetic case is
\begin{equation} \label{gauge1}
F = -q dx \wedge d\varphi,
\end{equation}
and the gauge field in the electric case is
\begin{equation} \label{gauge2}
F = -q dt \wedge dy,
\end{equation}
where $q=\sqrt{r_+ r_-}$. The roots of $G(x)$ in ascending order will
be denoted as $x_1,x_2,x_3, x_4$; similarly the roots of
$H(y)$ in ascending order are $y_1,y_2, y_3, y_4$. 

A redefinition of the parameters 
$a$, $b$, $m$, $q$, $A$ and $\Lambda$ can be compensated for by
a 3-parameter linear coordinate transformation which therefore maps
any $C$ metric onto any other $C$ metric up to a constant
conformal transformation \cite{Cmetric}.   This freedom is typically
used to eliminate the linear term in $G$ and $H$ and to set $a=1$ and
$b=1$. Rather than using the entire 3-parameter freedom, I shall
use only 1 parameter to eliminate the linear term, leaving $a$ and
$b$ arbitrary, as has been assumed in (\ref{ccmet}).  The parameter
$m$ will be assumed to be positive.

Under the
coordinate transformation $y=x-1/Ar$, $t=Au-\int^y dz/H(z)$, the metric 
(\ref{ccmet}) may also be written as
\begin{equation} \label{ccmetu}
ds^2 =  H(x-\frac{1}{Ar})A^2r^2 du^2 - 2dudr -2Ar^2dudx  +
r^2\left(G^{-1}(x) dx^2 + G(x) d\varphi^2 \right) \quad  .
\end{equation}
The electric field becomes 
\begin{equation}\label{gauge3}
F_E = -q du \wedge (Adx +\frac{dr}{r^2})
\end{equation}
and the magnetic field is unchanged.

Before considering the elimination of conical singularities in
(\ref{ccmet}), I shall first consider the structure of the polynomials
$G$ and $H$.  In order that the metric (\ref{ccmetu}) (or alternatively
(\ref{ccmet})) be 
of the appropriate signature, $G(x)$ must be positive over the allowed
range of $x$. If there are no degenerate roots of $H(y)$, then either
(a) $y<x$ or (b) $y>x$.
When the parameters in the $C$ metric are such that all roots of
$H$ are real, in case (a) the roots of $H$ in ascending order 
(i.e. $y_1,y_2,y_3$) are respectively the inner black hole horizon,
the outer black hole horizon, and the acceleration or cosmological horizon
respectively, the largest root having no physical meaning.
In case (b) the situation is reversed: the roots of $H$
in desending order (i.e. $y_4,y_3,y_2$) are respectively interpreted as
the inner black hole horizon,
the outer black hole horizon, and the acceleration or cosmological horizon,
the smallest root having no physical meaning.

The function $G$ has at most three extrema. These are located at
\begin{equation}\label{extrema}
x_m = {\displaystyle 
\frac{ - 3\,m - \sqrt{9\,m^{2} - 8\,q^{2}\,b}}{4q^{2}\,A}}, \,
x_0=0, \,
x_p = {\displaystyle 
\frac{ - 3\,m + \sqrt{9\,m^{2} - 8\,q^{2}\,b}}{4q^{2}\,A}} , 
\end{equation}
in ascending order if both $A>0$, and $b<0$ (i.e. $x_m<x_0<x_p$).
However if $A>0$ and $b>0$ then $x_m<x_p<x_0$.  
If $A<0$ then both these inequalities are reversed.

Consider next the elimination of conical singularities in the $(x,\varphi)$ sector.
For definiteness,  take $A>0$.  In order to have a
regular solution, there must be no conical singularities at either of
the endpoints of the domain of $x$, a criterion which can be satisfied in three
ways. 

(I) If the endpoints are a finite proper distance apart, then
this criterion becomes\cite{dgkt}
\begin{equation} \label{nostrut}
G'(x_i) =- G'(x_{i+1}),
\end{equation}
with $\varphi$ periodically identified with period $\Delta \varphi =
4\pi/ |G'(x_i)|$. Eq. (\ref{nostrut}) is equivalent to the condition
\begin{equation}\label{nostrut2}
\prod_{j\neq i}|(x_i-x_j)| = \prod_{j\neq i+1}|(x_i-x_j)|
\end{equation}
which can be satisfied by taking $x_i=x_{i+1}$ for $i=1,2,3$. For
$i=1$ or $i=3$, this is the only solution to (\ref{nostrut2}). For
$i=2$, a non-trivial solution (with $x_i \neq x_{i+1}$) also exists, but would yield a 
metric which does not have the correct signature. Consequently the only solution
to (\ref{nostrut}) is obtained  by setting any adjacent pair of roots equal.
This implies that $G(x)$ has an extremum at its (double) root.  The conformal
prefactor in front of the metric will therefore diverge only when $y=x_i=x_{i+1}$,
which can be taken to be the upper or lower limit of the range of $y$ as appropriate.

(II) There would be no conical singularity at $x=x_3$ if this point
were an infinite proper distance from any allowed value of $x$.
This condition requires that $x_2=x_3$, implying that $G(x)$
has an extremum (a minimum) at its middle (double) root. In this case
$\varphi$ is still periodically identified with period 
$4\pi/ |G'(x_4)|$, and the range of $x$ is $x_2=x_3 < x \leq x_4$.
The $(x,\varphi)$ sector is then no 
longer compact.  The conformal prefactor will diverge unless the range of
$y$ is restricted to be less than $x_2=x_3$. (Alternatively, one could choose
$ x_1 \leq x < x_2=x_3$ in which case $ y > x_2=x_3$). Signature requirements
then imply that $H(z) \leq G(z)$, {\it i.e.} the cosmological constant is negative
or zero.

(III) Conical singularities could be eliminated if the middle pair of roots of
$G(x)$ were complex. Although it would appear that the range of $x$ is 
$x_1 \leq x \leq x_4$, in this case it is not possible to ensure finiteness of the 
conformal prefactor over the entire allowed range of $y$. However if
$H(y)$ has a degenerate pair of roots
at $y=y_D$, one could then perform a coordinate transformation on $y$ such that
the proper distance between these roots is finite. 
If such a degenerate pair exists, then $x_1 < y_D < x_4$ and 
the conformal prefactor in front of the metric will diverge at $x=y_D$. Avoiding
this divergence then entails restricting the range of $x$ to $y_D < x \leq x_4$
rendering the $(x,\varphi)$ sector noncompact as in the previous case.
Conical singularities are removed by periodically identifying $\varphi$ with period 
$4\pi/ |G'(x_4)|$. (Alternatively one can choose
$y_D > x \geq x_1$, in which case the period of $\varphi$ must be
$4\pi/ |G'(x_1)|$ to avoid conical singularities).  This case can only 
occur for negative cosmological constant.

Hence removal of conical singularities in the cosmological $C$ metric
implies that either $G(x)$ or $H(y)$ must have a double root.  If such double roots
are separated by a finite proper distance, the general analysis of such roots is 
carried out by considering the metric
\begin{equation}\label{Fpsi}
  d\ell^2 = \varepsilon \frac{dz^2}{F(z)} + F(z) d\Psi^2
\end{equation}
where $F(z) = -\prod_{i=1}^4(z-z_i)$ is a quartic polynomial (either $G(x)$ or $H(y)$) and 
$\Psi$ is either $t$ ($\varepsilon=-1$)
or  $\varphi$ ($\varepsilon=+1$).  In the limit that a pair of adjacent roots $(z_j,z_{j+1})$ are equal, 
set
\begin{equation}\label{rtlim}
z=z_j + \epsilon f(\lambda)   \qquad    z_{j+1} - z_j = \epsilon f(\hat{\lambda}) \equiv
    2 \epsilon \hat{\eta}
\end{equation}
and take the limit as $\epsilon \to 0$. Rescaling $\Psi = \psi/\epsilon$ yields
\begin{equation}\label{fpsi}
  d\ell^2 = -\varepsilon \frac{df^2}{\prod_{i\neq j,j+1}^4 (z_j-z_i)f(f- 2\hat{\eta})} 
- \prod_{i\neq j,j+1}^4 (z_j-z_i) f(f- 2\hat{\eta}) d\psi^2
\end{equation}
for the metric in this sector. Setting $f = \eta + \hat{\eta}/2$ yields
\begin{equation}\label{etapsi}
  d\ell^2 = \varepsilon\frac{d\eta^2}{\prod_{i\neq j,j+1}^4(z_j-z_i)(\hat{\eta}^2-\eta^2)}
+ \prod_{i\neq j,j+1}^4(z_j-z_i)  (\hat{\eta}^2-\eta^2) d\psi^2
\end{equation}
as the generic result.  If either the largest or smallest pairs of roots of $F(z)$
are degenerate then $\prod_{i\neq j,j+1}^4(z_j-z_i) > 0$, whereas if the middle two
roots are degenerate then $\prod_{i\neq j,j+1}^4(z_j-z_i) < 0$.  Signature requirements
will then dictate the range of $|\eta|$.

I consider next what kinds of spacetimes
the metric (\ref{ccmet}) describes for each of the above cases.

\subsection{Case I}

Consider for definiteness, the situation when the double root is at $x=0$.
From (\ref{Gdef}) and (\ref{extrema}) this implies that
$\tilde{a}=0$ or $A^2 = \frac{\Lambda}{3a}$. Note
that solutions exist for both signs of $\Lambda$  provided the
signs of $a$ and $\Lambda$ match.  The usual parameter
choice $a=1$ therefore eliminates all $\Lambda < 0$ solutions, and
is therefore unnecessarily restrictive.  Setting
$\tilde{a} = k\epsilon^2$, the double-root condition is satisfied
in the limit $\epsilon \to 0$. Writing $\varphi = \phi/\epsilon$
and $x = \epsilon f(\lambda)$, the roots $x_i = \epsilon f(\lambda_i)$ and
$x_{i+1} = \epsilon f(\lambda_{i+1})$ coincide 
as $\epsilon \to 0$ for each of $i=1,2,3$.  From (\ref{etapsi})
the $(x,\varphi)$ section of the metric becomes
\begin{equation}\label{xphisect}
\frac{dx^2}{G(x)} + G(x) d\varphi^2 =
\frac{df^2}{k-bf^2} + (k-bf^2) d\phi^2 
\end{equation}
in the limit $\epsilon \to 0$, apart from the conformal factor.
Note that $\phi$ has period $2\pi$.

The full metric is of the form 
\begin{equation}
ds^2 = \frac{1}{A^2 y^2} 
\left[ H(y) dt^2 - H^{-1}(y) dy^2 + d\Omega^2_b \right],
\end{equation}
which after making the further coordinate transformation 
$y = -\frac{1}{Ar}$, $t = AT$ becomes
\begin{equation}\label{rnbmet}
ds^2 = -V(r) dT^2 + \frac{dr^2}{V(r)} + r^2d\Omega^2_b  ,
\end{equation}
where
\begin{equation}\label{rnpot}
V(r) = - \frac{\Lambda}{3}r^2 + b -\frac{2m}{r} + \frac{q^2}{r^2}  .
\end{equation}
The range of $r$  is from $0$ to $\infty$.

Degeneracy of the largest (or smallest) two roots of $G(x)$ occurs if and only if $b>0$;
in this case the class of metrics 
obtained are of the Reissner-Nordstr\"om (anti)-de Sitter type, 
of mass $m$ and charge $q$.
However there is a surprise in that the parameter $b$ is completely arbitrary. 
A simple rescaling of 
parameters and coordinates allows it  to be set to unit magnitude 
without loss of generality if it is nonzero (as is the case for $k$). 
If the middle two roots are degenerate, then $b=-1$, and if the largest (or smallest) 
three  roots are degenerate then $b=0$.

In all, there are five possible regular spacetimes which result, 
depending upon
the relative signs of $b$, $k$ and $\Lambda$. 
These are characterized by
the two-dimensional metric $d\Omega^2_b$:
\begin{eqnarray}
&b=1, k=1& \quad d\Omega^2_b = d\theta^2 + \sin^2(\theta)d\phi^2  
\label{b1k1} \\
&b=0, k=1& \quad d\Omega^2_b = d\theta^2 + \theta^2 d\phi^2 
\label{b0k1} \\
&b=-1, k=1& \quad d\Omega^2_b = d\theta^2 + \cosh^2(\theta) d\phi^2  
\label{bm1k1}\\
&b=-1, k=-1& \quad d\Omega^2_b = d\theta^2 + \sinh^2(\theta) d\phi^2  
\label{bm1km1}\\
&b=-1, k=0& \quad d\Omega^2_b = d\theta^2 + e^{(2\theta)} d\phi^2  
\label{bm1k0}
\end{eqnarray}
where an obvious coordinate transformation $f=f(\theta)$ has been employed 
in each case.
The $b=1$ case is valid for both signs of $\Lambda$, and so
corresponds to two different spacetimes,  whereas signature
requirements in the other four cases imply $\Lambda<0$.
The values of $k$ given above are likewise
determined by signature requirements.

It is easily checked that each of these six spacetimes satisfies
the Einstein-Maxwell equations with cosmological constant.
The gauge field becomes 
\begin{equation}
F = -\frac{q}{r^2} dT \wedge dr
\end{equation}
for all values of $b$ in the electric case, and
\begin{eqnarray}
&b=1, k=1& \quad  F = q\sin\theta d\theta \wedge d\phi 
\label{b1k1m} \\
&b=0, k=1& \quad F = q d\theta \wedge d\phi 
\label{b0k1m}\\
&b=-1, k=1& \quad F = q\cosh\theta d\theta \wedge d\phi 
\label{bm1k1m}\\
&b=-1, k=-1& \quad F = q\sinh\theta d\theta \wedge d\phi 
\label{bm1km1m}\\
&b=-1, k=0& \quad F = q e^\theta d\theta \wedge d\phi 
\label{bm1k0m}
\end{eqnarray}
in the magnetic case.

The preceding analysis holds in the case that the degenerate root is at $x=x_0=0$.
It is straightforward to check that this analysis does not qualitatively change if
the degenerate root is at either $x=x_m$ or $x=x_p$, and/or if the sign of $A$
is reversed and/or if the roles of the largest and smallest roots are reversed,
although the intermediate steps differ, as does the relationship between $a$, $A$
and $\Lambda$. In other words, these six spacetimes are the only regular ones
that can result from the requirement that the cosmological C-metric (\ref{ccmet})
be free of conical singularities  for case I.

The regularity requirements for the $C$ metric in
the $b=1$ case for positive $\Lambda$ have been discussed 
previously in ref. \cite{cosern}. The other five spacetimes, however, 
have been overlooked in previous studies.  The $b=1$, $\Lambda<0$ case is simply
Reissner-Nordstr\"om anti-de Sitter spacetime.  The remaining
three spacetimes all have $\Lambda <0$ and $b<0$, and necessarily have
non-trivial topology, as I will demonstrate in the next section.

\subsection{Case II}

In this case the middle two roots of  $G(x)$ are degenerate, and $x$ ranges
from its largest (smallest) root to this central degenerate root which is an 
infinite proper distance away. This implies that
\begin{equation}\label{Gdegen}
G(x) =  - q^{2}\,A^{2}\,(x - x_+)\,(x - x_0)^{2} \,(x - x_-)
\end{equation}
must hold, where by definition $x_+ > x_0 > x_-$. From (\ref{Gdef}),  there
can be no term linear in $x$ in the right-hand-side of the above equation, which
implies
\begin{equation}\label{midrts}
{x_0} = 0, \, - 2\,{\displaystyle \frac {{x_-}\,{x_+}
}{{x_+} + {x_-}}} 
\end{equation}
and so there are two possible values for the central root.  
Note that for $\Lambda>0$, $G(z)<H(z)$, so when the central
two roots of $G(x)$ are degenerate $H(y)$ has only two roots $y_-$ and $y_+$.
For $y_- < y < y_+$, $y$ is a timelike coordinate, whereas for $y$ outside
these bounds the metric has a naked curvature singularity. Hence it is not
possible to keep $\Lambda$ positive and maintain the correct metric signature
requirements without avoiding naked singularities.

If $x_0=0$ is the central root, then $0 < {x_+}$ and  ${x_-} < 0$. 
For $A>0$, a calculation then demonstrates that 
\begin{equation}\label{xlxm0}
x_\pm = \pm \frac {\sqrt{m^{2} - q^{2}\,b} \mp m}{q^{2}\,A}
\end{equation}
and the preceding inequalities hold provided $b<0$ and 
$A^2 = \frac{\Lambda}{3a}$; the roles of $x_\pm$ are interchanged if $A$
reverses sign.  After a coordinate transformation
$x\to 1/(Ar)$ $y \to 1/(AR)$ $t \to AT$ and $\varphi \to \frac{2Ar_+^2}{r_+-r_-}\phi$,
the resultant metric is 
\begin{eqnarray}\label{cmetlR1}
\lefteqn{{ds}^{2}= - \frac{(R^{2}/l^2 - 1 - 2\frac{m}{R} + \frac{q^{2}}{R^2})\,r^{2}\,{dT^2}}
{(R + r)^{2}}  
+ \frac {r^{2}{dR^2}}{(R^{2}/l^2 - 1 - 2\frac{m}{R} + \frac{q^{2}}{R^2})\,(R + r)^{2} } } 
\nonumber \\
 & & \mbox{} + {\displaystyle \frac {r^{2}\,R^{2}\,{\it \ d}\,r^{
{\it 2\ }}}{(r^{2} - 2\,m\,r - q^{2})\,(R + r)^{2}}}  + 
{\displaystyle 
\frac {4 r_+^2 (r^{2} - 2\,m\,r - q^{2})\,R^{2}\, d \phi^2}{(r_+ - r_-)^2 r^{2}\,(R + r)^{2}}} 
\mbox{\hspace{142pt}}
\end{eqnarray}
where $\Lambda = -3/l^2 < 0$ and $r_\pm = m \pm \sqrt{m^2+q^2}$.
The metric (\ref{cmetlR1}) is free of conical singularities in the 
$(r,\phi)$ section provided $\phi$ has period $2\pi$. It satisfies the Einstein-Maxwell 
equations, where in the electric case
\begin{equation} \label{gauge1Re}
F = -\frac{q}{R^2} dT \wedge dR,
\end{equation}
and
\begin{equation} \label{gauge1Rm}
F = -\frac{q}{r^2} dr \wedge d\phi,
\end{equation}
in the magnetic case. 

If  $x_0 =\frac {{x_-}\,{x_+}}{{x_+} + {x_-}}$ is the central root,
then $x_+>0$ and either ${x_+} <  - 3\,{ x_-}$ or 
$ - \frac {{x_+}}{3} < {x_-} < 0$.  The analysis of (\ref{Gdegen}) is again
straightforward (although somewhat more complicated) and yields (for $A>0$)

\begin{equation}\label{xlxm2}
x_\pm = \pm \frac {m\,(2\,\sqrt{1 + \sqrt{9 - \frac {8\,b\,q^{2}}{m^{2}}}} 
\mp (1 + \sqrt{9-\frac {8\,b\,q^{2}}{m^{2}}}))}{4\,q^{2
}\,A}
\end{equation}
and
\begin{equation}\label{xm2}
x_0 = \frac {m\,( - 3 + \sqrt{9-\frac {8\,b\,q^{2}}{m^{2}}})}{4\,q^{2}\,A}
\end{equation}
provided
\begin{equation}\label{aaterm}
a = \frac{\Lambda}{3A^2} 
- \frac {27\,m^{4} - 36\,b\,q^{2}\,m^{2} + 8\,b^{2}\,q
^{4}}{32\,q^{6}\,A^{2}} - \frac {m\,\sqrt{(9\,m^{2} - 8\,q^{2}\,b
)^{3}}}{32\,q^{6}\,A^{2}}
\end{equation}
and $b>0$. As a consequence $0 \leq q^{2} \leq \frac {9\,m^{2}}{8}$. The
metric is
\begin{eqnarray}\label{cmetlR2}
\lefteqn{{\it ds}^{2}= - {\displaystyle \frac {\hat{r}^{2}\, \left( 
 \! {\displaystyle \frac {\hat{R}^{2}}{l^2}}  - {\displaystyle 
\frac {(3\sqrt{g}-g)\,m^{2}}{4q^{2}}}  + {\displaystyle \frac {(1- 
\sqrt{g})\,m}{\hat{R}}}  + {\displaystyle \frac {q^{2}}{\hat{R}^{2}}} 
 \!  \right) \, d\hat{T}^{2}}{(\hat{R} + \hat{r})^{2}}} } 
\nonumber \\
 & & \mbox{} + {\displaystyle 
\frac{\hat{r}^{2}\,{\it d\hat{R}}^{2}}{(\hat{R} + \hat{r})^{2}\, 
\left(  \! {\displaystyle \frac {\hat{R}^{2}}{l^2}}  -
{\displaystyle \frac {(3\sqrt{g}-g)\,m^{2}}{4q^{2}}}  + 
{\displaystyle \frac {( 1- \sqrt{g})\,m}{\hat{R}}}  + 
{\displaystyle \frac {q^{2}}{\hat{R}^{2}}}  \!  \right) }}  \\
 & & \mbox{} + {\displaystyle \frac {\hat{r}^{2}\,\hat{R}^{2}\,{\it d\hat{r}}^{2}}{
 \left(  \! {\displaystyle \frac {(3\sqrt{g}-g)\,m^{2}\,\hat{r}^{2}}{4q^{2}}}  
+ ( 1- \sqrt{g})\,m\,\hat{r} - q^{2} \!  \right) \,(\hat{R} + \hat{r})^{2}}}  \\
 & & \mbox{} + {\displaystyle 
\frac {\hat{R}^{2}\,((3\,\sqrt{g}-g)\,\frac{m^{2}}{4q^2}\,\hat{r}^{2} + 
(1-\sqrt{g})m\,\hat{r} - \,q^{2})\, d\hat{\phi}^{2}}{r^{2}\,(\hat{R} + \hat{r})^{2}}} 
\end{eqnarray}
where $g\equiv  9-\frac{8\,b\,q^{2}}{m^{2}}$. 
Since $0<g<9$, the term $\lambda^2 \equiv \frac {(3\sqrt{g}-g)\,m^{2}}{4q^{2}} > 0$.
Rescaling the coordinates $(\hat{T},\hat{R},\hat{r},\hat{\phi})\rightarrow 
(T/\lambda,\lambda R, \lambda r, \tilde{\phi}/\lambda)$ then yields
\begin{eqnarray}\label{cmetlR2a}
\lefteqn{{\it ds}^{2}= - {\displaystyle \frac {r^{2}\, \left( 
 \! {\displaystyle \frac { R^{2}}{l^2}}  - 1 
+ {\displaystyle \frac {(1- \sqrt{g})\,m}{\lambda^3 R}}  
+ {\displaystyle \frac {q^{2}}{\lambda^4 R^{2}}} 
 \!  \right) \, dT^{2}}{(R + r)^{2}}} } \nonumber\\
 & & \mbox{} + {\displaystyle 
\frac{r^{2}\,{\it dR}^{2}}{(R + r)^{2}\, 
\left(  \! {\displaystyle \frac { R^{2}}{l^2}}  - 1
+ {\displaystyle \frac{( 1- \sqrt{g})\,m}{\lambda^3 R}}  + 
{\displaystyle \frac {q^{2}}{\lambda^4 R^{2}}}  \!  \right) }}  \nonumber\\
 & & \mbox{} + {\displaystyle \frac {r^{2}\,R^{2}\,{\it dr}^{2}}{
 \left(  \! r^2 + \frac{( 1- \sqrt{g})\,m}{\lambda^3}r
- \frac{q^{2}}{\lambda^4} \!  \right) \,(R + r)^{2}}}  
+ {\displaystyle 
\frac {R^{2} (r^{2} + \frac{( 1- \sqrt{g})\,m}{\lambda^3}r - \frac{q^{2}}{\lambda^4}) 
d\tilde{\phi}^{2}}{r^{2}\,(R + r)^{2}}} 
\end{eqnarray}
which is qualitatively the same as the metric (\ref{cmetlR1}) once $m$ and $q$ are
rescaled.  However the coefficient $(1-\sqrt{g})$ multiplying $m$ can be either 
positive or negative.  

Hence the general form of the case II metric is
\begin{eqnarray}\label{caseII}
\lefteqn{{ds}^{2}= - \frac{(R^{2}/l^2 - 1 - 2\,M/R + Q^{2}/R^2)\,r^{2}\,{dT^2}}
{(R + r)^{2}}  
+ \frac {r^{2}{dR^2}}{(R^{2}/l^2 - 1 - 2\,M/R + Q^{2}/R^2)\,(R + r)^{2} } } \nonumber\\
 & & \mbox{} + {\displaystyle \frac {r^{2}\,R^{2}\,{\it \ d}\,r^{
{\it 2\ }}}{(r^{2} - 2\,M\,r - Q^{2})\,(R + r)^{2}}}  + 
{\displaystyle 
\frac {4 r_+^2 (r^{2} - 2\,M\,r - Q^{2})\,R^{2}\, d \phi^2}{(r_+ - r_-)^2 r^{2}\,(R + r)^{2}}} 
\mbox{\hspace{142pt}}
\end{eqnarray}
where $m$ and $q$ have been appropriately rescaled, and $\phi$ has been rescaled so
as to removed spurious conical singularities.  The parameter $M$ may have either sign.
The electromagnetic field strength tensors are the same as (\ref{gauge1Re}) and 
(\ref{gauge1Rm}) respectively with $q\to Q$. 

The coordinate $R$ has the range $0 \leq R \leq \infty$ whereas
signature requirements demand that the coordinate $r$ have the range 
$r_+ \leq r \leq \infty$, where $r_{\pm} = M \pm \sqrt{M^2+Q^2}$. 
reminiscent of the type I instanton discussed in ref. \cite{cosern}. For 
large $r$ the metric (\ref{cmetlR1}) 
asymptotically approaches the metric (\ref{rnbmet}) where $d\Omega_b^2$ is 
given by (\ref{bm1k0}). Near $r=r_+$ the metric is conformal to (\ref{rnbmet}),
where $d\Omega_b^2$ is given by (\ref{b0k1}).  There is a curvature singularity
at $R=0$, and event horizons at the roots of the equation $R^2/l^2-1-2M/R +Q^2/R^2=0$.
These horizons are not compact surfaces.

\subsection{Case III}

This case is similar to case I, but with the roles of $(x,\varphi)$ and $(y,t)$
reversed.  

Again, suppose for definiteness that $H(y)$ has a double root at $y=0$.
From (\ref{Hdef}) and (\ref{extrema}) this implies that ${a}=0$.  Setting 
${a} = k\epsilon^2$, and writing $T = t/\epsilon$
and $y = \epsilon R(\lambda)$, the roots $y_i = \epsilon R(\lambda_i)$ and
$y_{i+1} = \epsilon R(\lambda_{i+1})$ coincide 
as $\epsilon \to 0$ for each of $i=1,2,3$.  From (\ref{etapsi})
the $(t,y)$ section of the metric becomes
\begin{equation}\label{tysect}
-\frac{dy^2}{H(y)} + H(y) dy^2  =
- (bR^2-k) dT^2 + \frac{dR^2}{bR^2-k} 
\end{equation}
as $\epsilon \to 0$, apart from the conformal factor. Signature requirements
permit all possible signs for $b$ and $k$, although for $b<0$ and $k>0$ 
$R$ is a timelike coordinate and the metric is no longer static. 

After making the coordinate transformation $x = -\frac{1}{Ar}$, 
$\varphi = A\hat{\phi}$ the full metric becomes
\begin{equation}\label{cas3met}
ds^2 = r^2\left( -(bR^2-k) dT^2 + \frac{dR^2}{bR^2-k}\right)
+ \frac{dr^2}{U(r)} + U(r) d\hat{\phi}^2
\end{equation}
where
\begin{equation}
U(r) = \frac{|\Lambda|}{3}r^2 - b + \frac{2m}{r} - \frac{q^2}{r^2}  .
\end{equation}
Provided that either $q^2 > \frac{b^2l^2}{12}$ or
$m > m_+\equiv \frac{\hat{r}_+}{2}(\frac{\hat{r}^2_+}{l^2}+\frac{q^2}{\hat{r}^2_+})$
or
$m < m_-\equiv \frac{\hat{r}_-}{2}(\frac{\hat{r}^2_-}{l^2}+\frac{q^2}{\hat{r}^2_-})$
where
\begin{equation}\label{rhatdef}
\hat{r}^2_\pm = {\displaystyle \frac {1}{6}} \,l^{2}\,b\,(1 \pm \sqrt{1
 - 12\,{\displaystyle \frac {q^{2}}{b^{2}\,l^{2}}} })
\end{equation}
the function $U(r)$ will have only one positive root $r=r_U$, and the range of $r$ is   
$r_U < r < \infty$.   If $m_- < m < m_+$ then  $U(r)$ will have three positive roots,
and the range of $r$  will be $r_M < r < \infty$ where $r_U$ is now 
the largest root of $U(r)$.  In both cases conical singularities are not present 
provided $\hat{\phi}$ has period $\frac{4\pi}{|U'(r_U)|}$. 
Although signature requirements also permit $r$ to lie between the smallest two positive
roots of $U(r)$, it is not possible to eliminate conical singularities in the 
$(r,\hat{\phi})$ section.  

The magnetic gauge field (\ref{gauge1}) becomes
\begin{equation} \label{c3g1}
F = -\frac{q}{r^2} dr \wedge d\hat{\phi},
\end{equation}
which is like an electric field, whereas the electric gauge field (\ref{gauge2})
is now
\begin{equation} \label{c3g2}
F = -qdR \wedge dT,
\end{equation}
is like a magnetic field where, for example, $R=\cos(\chi)$ when $k=b=-1$, yielding
$F = q \sin(\chi) d\chi \wedge dT$.  

The class of metrics (\ref{cas3met}) are  products of a 2 dimensional
(anti) de Sitter spacetime (which is a black hole for $b>0$ and $k>0$ \cite{dan}) 
with a Euclidean 2 dimensional de Sitter space.  Employing the coordinate
transformation $\rho^2 = U(r)l^2$, these metrics become
\begin{equation}\label{cas3rho}
ds^2 = [U^{(-1)}(\rho^2/l^2)]\left( -(bR^2-k) dT^2 + \frac{dR^2}{bR^2-k}\right)
+ \frac{d\rho^2}{U'(U^{(-1)}(\rho^2/l^2))} + \rho^2 d\phi^2
\end{equation}
where $\hat{\phi} =2l|U'(r_U)|\phi$ and $U^{(-1)}$ is the inverse of $U$,
i.e. $U(U^{(-1)}(z)) = z$.  For large $r$, $\rho\approx r$, and the metric
and these metrics are all asymptotic to 
\begin{equation}\label{cas3ban}
ds^2 = ({\rho^2}+b l^2)\left( -(bR^2-k) dT^2 + \frac{dR^2}{bR^2-k}\right)
+ \frac{d\rho^2}{\frac{\rho^2}{l^2}+b} + \rho^2 d\phi^2
\end{equation}
which is the product of $(2+1)$ (anti) de Sitter spacetime and a circle. When
$m=q=0$ the metric (\ref{cas3ban}) is exactly equal to (\ref{cas3met}) after
the coordinate transformation $\rho^2 = r^2-bl^2$.
For $b<0$  these are $(3+1)$ dimensional versions of the constant 
curvature black holes recently considered by Banados \cite{Bancon}.
If either of $m$ or $q$ are non-zero, the metric (\ref{cas3rho}) has 
naked singularities. 

If the degenerate roots of $H(y)$ are not at $y=0$, the situation is similar to the
situation just described, and the resultant class of metrics is still given
by (\ref{cas3met}), but with $\Lambda$, $m$ and $b$ redefined. However
the analysis is somewhat more complicated and will not be reproduced here.

\section{Topological Anti de Sitter Black Holes}
\label{tads}

The case I metrics (\ref{rnbmet})
yield an interesting class of topological black holes which I shall describe
in this section.

The $b=1$ metrics (\ref{b1k1}) correspond to the usual Reissner-Nordstrom de
Sitter and Reissner-Nordstrom anti de Sitter spacetimes respectively, depending
upon the sign of $\Lambda$.  The event horizons have the topology of a 2-sphere.

The remaining spacetimes all have $\Lambda <0$ and $b\leq 0$. Surfaces
of constant $T$ and $r$ (including the event horizon) are apparently 
noncompact spaces of constant nonpositive curvature. For $b=0$ the existence of 
such ``black plane'' spacetimes have recently been noted \cite{bplane}.  
However by appropriate identification of the coordinates it is possible to
render these surfaces compact for all $b\leq 0$.  Since they are also surfaces
of constant negative curvature they must have a non-trivial topology, which is
in turn inheirited by the entire spacetime.

Consider first $b=0$. The$(\theta,\phi)$ section is flat space. Geodesics in this
space are straight lines. By a trivial coordinate transformation the metric in
this section may be written as
\begin{equation}\label{lamtor}
d\Omega^2_b = d\lambda^2 + d\tilde{\phi}^2 
\end{equation}
where the geodesics are given by the equations $\alpha_1\lambda + \alpha_2\phi = \alpha_3$
where the $\alpha_i$ are constants.  Identifying $\lambda$ and $\tilde{\phi}$ 
with their own periodicities the $(\theta,\phi)$ sector becomes a torus. Its 
unit area can be chosen to be $4\pi$ by identifying the $\lambda$-coordinate with period 2
and the $\tilde{\phi}$ coordinate with period $2\pi$.

The remaining spacetimes (\ref{bm1k1}), (\ref{bm1km1}), (\ref{bm1k0}) all have $b=-1$. 
The $(\theta,\phi)$ sections are non-compact and respectively have
the topologies of a tube flared out at both ends, a two-sheeted hyperboloid,
and a tube flared at one end.  

Consider first the spacetime (\ref{bm1km1}), taking one sheet of the
hyperboloid. The $(\theta,\phi)$ section may be mapped to the Poincar\'e disk 
under the transformation
$\lambda = \tanh(\theta/2)$, yielding
\begin{equation}\label{pdisk}
d\theta^2 + \sinh^2(\theta)d\phi^2
= \frac{1}{(1-\lambda^2)^2}\left(d\lambda^2 + \lambda^2 d\phi^2\right)
\end{equation}
where $ 0\leq \lambda < 1$.  The Poincar\'e disk has an
isomorphism group SO(2,1). Identifying points on
the disk under any discrete subgroup of SO(2,1) yields a compact 
two-dimensional space of negative curvature, which necessarily has
genus $g \geq 2$. The unit area of such surfaces is $4\pi(g-1)$.
These spaces may be constructed by symmetrically placing a polygon of $4g$ sides
at the center of the Poincare disk and identifying opposite sides. 
The edges of the polygon are geodesics of the Poincar\'e disk; these are circles
whose extensions are orthogonal to the disk boundary. The simplest
case is the octagon with $g=2$.
The $q=m=0$ versions of these spacetimes can be understood as 
four-dimensional analogues of the three-dimensional black hole
\cite{btz}, as shown recently by Aminneborg {\it et.al.} \cite{amin}.  

Carrying out an analogous procedure on the other two spacetimes
(\ref{bm1k0}) and (\ref{bm1km1}) yields nothing new. This is because
each of their $(\theta,\phi)$ sections can locally be mapped into the
$(\theta,\phi)$ section of  (\ref{bm1km1}).  For example the local
transformation
\begin{equation}\label{tk1tokm1}
\sinh(\sigma) = \cos(\phi) \sinh(\theta) \qquad
\tanh(\Phi) = \sin(\phi) \tanh(\theta)
\end{equation}
yields
\begin{equation}\label{k1tokm1}
d\sigma^2 + \cosh^2(\sigma) d\Phi^2 \to d\theta^2 + \sinh^2(\theta) d\phi^2
\end{equation}
whereas
\begin{eqnarray}\label{tk0tokm1}
{ \chi} &=& {\displaystyle \frac {{\rm sinh}(\,{ \theta}\,)\,{\rm 
sin}(\,{ \phi}\,)}{{\rm cosh}(\,{ \theta}\,) + {\rm sinh}(\,{ 
\theta}\,)\,{\rm cos}(\,{ \phi}\,)}}
\nonumber\\
{ \alpha} &=& {\rm ln}(\,{\rm cosh}(\,{ \theta}\,) + {\rm sinh}(\,
{ \theta}\,)\,{\rm cos}(\,{ \phi}\,)\,)
\end{eqnarray}
yields
\begin{equation}\label{k0tokm1}
d\alpha^2 + \exp(2\alpha) d\chi^2 \to d\theta^2 + \sinh^2(\theta) d\phi^2 \qquad .
\end{equation}
Hence any local region in the $(\theta,\phi)$ section of (\ref{bm1km1}) containing
the $4g$-sided polygon can, along with the polygon, be mapped into a local region
of either of (\ref{bm1k1}) or (\ref{bm1k0}). The identification procedure then
follows through. Similarly, The electromagnetic field tensors 
(\ref{bm1k1m}) and (\ref{bm1k0m}) correspondingly map into (\ref{bm1km1m}).
 Hence without loss of generality the spacetimes (\ref{bm1k1})
and (\ref{bm1k0}) may be dropped from further consideration.
These $b\leq 0$ 
constructions hold for all values of $r$ and $T$ in (\ref{rnbmet}).

For $\Lambda < 0$, the metric function $V(r)$ in (\ref{rnpot}) has no term linear 
in $r$, and  the product of its roots is equal to $q^2l^2$. Hence
$V(r)$ has at most two roots for positive $r$ corresponding to an inner
and outer horizon, as with the usual RNadS metric. For $b=0$,
provided 
\begin{equation}\label{b0hor}
27\,l^{2}\,{\it m}^{4} \geq 16\,q^{6}
\end{equation}
there will be two horizons, 
with the extremal case saturating the inequality. For nonzero $b$
event horizons exist provided
\begin{equation}\label{bhor}
m^2 \leq 
  {\displaystyle \frac {l^2}{27}} \,{\displaystyle \frac { 16 -
24\,e^{2}b -16b\sqrt{1 - e^{2}b}\,e^{2} + 6b^2\,e^{4} 
+ 16\,\sqrt{1 - e^{2}b}}{e^{6}}} 
\end{equation}
where $e = \frac{2\sqrt{2}q}{3m}$. If $b=1$ (the genus 0 case)
a necessary condition for
event horizons to exist is that $q<m$, since for $q>m$ the right-hand side
of (\ref{bhor}) becomes negative and so (\ref{bhor}) cannot be satisfied;
the range of e is therefore from 0 to $2\sqrt{2}/3$. If $b=-1$ then there is no
(obvious) upper limit on e, and event horizons can exist for arbitrarily
large values of $q$ relative to $m$.

Hence for all values of $b$, the topology of
the outer event horizon is $H^2_g$, where $H^2_g$ is a two-dimensional
surface of genus $g$, with $g=0$ being the 2-sphere. The entire spacetime 
has topology $R^2\times H^2_g$.

An analysis of the evaluation of the mass and charge of the metrics given
by (\ref{rnbmet}) may be carried out using the quasilocal formalism developed
for anti de Sitter spacetimes \cite{bjm}.  Consider
a surface $B_g$ of topology $H_g$ at a fixed value $R$ of
the coordinate $r$ centered about the origin.  Since $\partial/\partial t$ is
a surface-forming Killing vector proportional to the normal of $B_g$, the conserved
mass parameter is given by 
\begin{equation}\label{quasimass}
M = \int_{B_g} d\Omega_g \sqrt{V(R)}\left(\frac{k}{\kappa}-\epsilon_0\right)
\end{equation}
where $\kappa = 8\pi$ is the gravitational coupling constant and $k$ is the trace
of the extrinsic curvature of the surface $B_g$ considered as a boundary $\partial\Sigma_g$
of a spacelike hypersurface $\Sigma_g$ whose unit normal is orthogonal to the normal
of $B_g$.  The quantity $\epsilon_0$ is the energy density associated with some
reference spacetime. Although it is not unique, in order to make a meaningful comparison
for a given topology, the reference spacetime should be chosen to be a spacetime
with the same topology as the original spacetime.  The natural choice would be
a spacetime with $m=0=q$ -- these are the massless AdS black holes considered
in refs. \cite{wendy,amin}. The trace of the extrinsic curvature for the boundary 
for $\Lambda = -3/l^2$ is
\begin{equation}
k = -\frac{2}{R}\sqrt{R^2/l^2+b-2m/R+q^2/R^2}
\end{equation}
independent of the topology. Taking $\epsilon_0$ to be equal to $k$ when $m=0=q$ yields
in the limit of large $R$
\begin{equation}\label{qmass}
M = \frac{m}{4\pi} \int_{B_g} d\Omega_g =  m(|g-1|+\delta_{g,1})
\end{equation}
as the conserved ADM mass of the spacetime. A similar analysis of the charge $Q$ 
contained within the same boundary $B_g$ indicates that 
$Q = q(|g-1|+\delta_{g,1})$ is the conserved charge of the black hole.

It is straightforward to show \cite{bjm} that the quantity $M$ in (\ref{quasimass}) is 
simply the Hamiltonian derived from the action 
\begin{equation}\label{actquasi}
S = -\frac{1}{2\kappa} \int
d^4x\sqrt{g}\left(R-2\Lambda-F_{\mu\nu}F^{\mu\nu}\right) +
\frac{1}{\kappa} \int^{\Sigma_f}_{\Sigma_i} d^3x \sqrt{h} K,
-\frac{1}{\kappa} \int_{\cal T} d^3x \sqrt{\gamma} \Theta  - S_0
\end{equation} 
evaluated when the constraints hold, in the large $R$ limit.
Here $\kappa = 8\pi$, and ${\cal T} =  B \times I$ is a timelike hypersurface 
(with induced metric $\gamma_{ij}$ and extrinsic curvature $\Theta_{ij}$)
joining the initial and final hypersurfaces
${\Sigma_i}$ and ${\Sigma_f}$ (with induced metric $h_{ij}$ extrinsic 
curvature $K_{ij}$) respectively.
$S_0$ is the reference action which yields $\epsilon_0$, and is a functional
of the metric on the boundary.

\section{Topological Black Hole Instantons}
\label{instsec}

From the preceding sections it is clear that the 
only special cases of the cosmological $C$ metrics for which
the metric is regular and the event horizon is compact reduce either to one 
of the Reissner-Nordstr\"om 
(anti) de Sitter metrics (\ref{rnbmet}) or the constant curvature black holes
(\ref{cas3ban}).   A consideration of the pair creation of black holes then 
reduces to a consideration of the non-singular instantons that can be 
constructed from either of these cases. The remaining metrics have either 
naked singularities  (\ref{cas3rho}) or non-compact event horizons 
(\ref{caseII}).  The possibility of compactifying these horizons will 
not be considered here.

For the Reissner-Nordstrom (anti) de Sitter (RN(a)dS) spacetimes the general
form of the metric can be written as
\begin{equation}\label{rnbmet2}
ds^2 = -N(r)dt^2 + \frac{dr^2}{N(r)} 
+ r^2d\Omega^2_b  ,
\end{equation}
where
\begin{equation}\label{Ndef}
N(r) = -s\frac{r^2}{l^2} + b -\frac{2m}{r} + \frac{q^2}{r^2}
\end{equation}
with $l^2 = \frac{3}{|\Lambda|}$, $s=\frac{|\Lambda|}{\Lambda}$ is the sign of
$\Lambda$, and
\begin{equation}\label{topsurf}
d\Omega_b = \left\{ \begin{array}{ll}
                     d\theta^2 + \sin^2(\theta)d\phi^2 & b=1, s=\pm 1 \\
                     d\theta^2 +  d\phi^2 & b=0, s= - 1 \\
                     d\theta^2 + \sinh^2(\theta)d\phi^2 & b=-1, s= - 1
              \end{array} \right\}
\end{equation}
corresponding to the genus $g=0$, $g=1$ and $g\geq 2$ cases respectively. 
The range of $r$  is from $0$ to $\infty$.  The gauge field is
\begin{equation}
F = -\frac{q}{r^2} dt \wedge dr \label{el1}
\end{equation}
for an electrically-charged solution, and 
\begin{equation}\label{topsurf2}
F = \left\{ \begin{array}{ll}
                     q \sin(\theta)d\theta\wedge d\phi & b=1, s=\pm 1 \\
                     q d\theta\wedge d\phi & b=1, s=-1 \\
                     q  \sinh(\theta)d\theta\wedge d\phi & b=1, s=-1 \\

              \end{array} \right\}
\end{equation}
in the magnetic case.

I shall restrict my attention to magnetically charged black holes, and
later consider electrically charged ones.
Instantons can be constructed from the metric (\ref{rnbmet2}) by analytically
continuing $t \to i\tau$. In order to obtain a positive-definite metric
the coordinate $r$ must lie in a region that ensures $N(r) > 0$. 
For values of the coordinate $r\geq r_H$ (where $r_H$ is the outer horizon)
such that $N(r_H)=0$ there is potentially
a conical singularity at this point. If $N^\prime(r_H)\neq 0$, then
a necessary condition for the regularity of the instanton is that
this conical singularity be removed by making $\tau$ periodic with
period 
\begin{equation}\label{betaH}
\beta_H=\frac{2\pi}{\kappa_H }= \frac{4\pi r_H}{-\frac{3sr^2_H}{l^2}
+{b}-\frac{q^2}{r^2_H}}
\end{equation}
where $\kappa_H$ is the surface gravity at $r=r_H$.
If $\kappa_H=0$ then this point is an infinite proper distance from any 
other point, and so  $r > r_H$ and $\tau$ can be identified with arbitrary 
period \cite{entar}.

Since $N(r)$ diverges for large $r$, these instantons will only be regular
provided one of the following additional conditions is satisfied.
\begin{itemize}
\item[(A)]If $N \to -\infty$ then $N(r)$ must have another 
root $r=r_C > r_H$, interpreted as a cosmological horizon. 
In this case a regular instanton can be obtained by either
(i) identifying $\tau$ with period $2\pi/\kappa_C$ (when $\kappa_H = 0$) or
(ii) setting $|\kappa_C| = |\kappa_H|$ (when $N^\prime(r_H)\neq 0$) so that the 
periodicity at both horizons is the same. 
\item[(B)]If $N \to \infty$ then there are no other roots of $N(r)$ for the metric
(\ref{rnbmet2}).  The instanton, although regular, is not compact, and must be
modified by including additional stress energy in order to ameliorate this difficulty.
\end{itemize}

Case (A) is the de Sitter case ($s=1$), and 
a discussion of its instantons has already appeared \cite{cosern}.
Here I briefly recapitulate the results. Since there are two horizons,
there are in all four types of instantons: 
\begin{itemize}
\item[(a)]lukewarm, when $|\kappa_C| = |\kappa_H|$ but $r_C\neq r_H$ (implying $q=m$) 
\item[(b)]charged Nariai, when $|\kappa_C| = |\kappa_H|$ and $r_C = r_H$ 
\item[(c)]cold, if $\kappa_H=0$ and $r_C\neq r_H$
\item[(d)]ultracold, if $\kappa_H=0$ and $r_C = r_H$
\end{itemize}
(There is a second ultracold instanton, but it does not have horizons \cite{cosern}).
For sufficiently small mass, both lukewarm and cold instantons can exist, 
which respectively correspond to pair creation of non-extreme and extreme 
black holes. At given mass, the
cold solution has higher charge than the lukewarm solution. 
For $m \geq l/(3\sqrt{3})$, the charged Nariai instanton is viable; it
has lower charge than the other two. For $m \geq 3l/4$, the lukewarm
and charged Nariai solutions coincide, and there is no lukewarm
solution with higher mass. The cold and charged Nariai solutions
coincide in the ultracold solution, when $m=2l/(3\sqrt{6})$, 
and there are no regular solutions where the mass is larger than this.

Case (B) is the anti de Sitter case. As previously noted \cite{cqgtop}
pair production of these black holes may be achieved using domain walls
\cite{dompair,Chamsugra}.  Since the gravitational field
of a domain wall is repulsive, inclusion of a domain wall into the
anti de Sitter case can provide the necessary energy to pair create the black
holes, analogous to the manner in which the positive cosmological constant performs
a similar function in the de Sitter case.  Of course there is nothing obstructing
the inclusion of domain walls in the de Sitter case, and it can be included in
the analysis as well.

The general situation involves constructing a two-sided bubble by taking two regions
of the RNadS spacetime and joining them together along a common timelike boundary which
is homeomorphic to $H^2_g\times R$.  The boundary along which they are joined must
satisfy the Israel matching conditions. In the Lorentzian section the result is a
spacetime with two domains, each of which contains a black hole.  The topology
of the RNadS Riemannian section is   $R^2\times H^2_g$, where the
$R^2$ factor is like a bell.  Two copies of this
manifold may be matched together at a radius $r$ at the open end of their bells
determined by the matching condition.  The resultant Riemannian section now has 
topology  $S^2\times H^2_g$
which contains a single domain wall of topology $S^1\times H^2_g$ and two bolts
of topology $H^2_g$ where the Killing field $\frac{\partial}{\partial \tau}$ vanishes.
The nucleation surface $\Sigma$ which joins the Lorentzian and Riemannian sections
is located along the $\tau=0$ and $\tau=\beta_H/2$ segments.

The matching conditon for the class of spacetimes given by (\ref{rnbmet2}) may
be obtained in a manner completely analogous to the genus zero cases \cite{His,Aur} 
and is given by
\begin{equation}\label{match}
\sqrt{N(r) - \dot{r}^2} = 2\pi\sigma r
\end{equation}
where $\sigma$ is the energy per unit area of the domain wall, whose
topology is $S^1\times H^2_g$, and the overdot refers to the derivative
with respect to Euclidean proper time. Equation (\ref{match}) may be interpreted
as the equation describing the motion of a fictitious particle
in a potential $v = N - (2\pi\sigma r)^2$. 

Static solutions for which $r=r_s$  have energy zero, and  may
be obtained by solving (\ref{match}) under the condition 
$\partial{v}/\partial{r} = 0$.  These are given by
\begin{equation}\label{matchr}
r^2_s =  \frac{l^2}{6\gamma}\left[ b\pm \sqrt{b^2-12\frac{q^2\gamma}{\l^2}}\right]
\end{equation}
where $\gamma \equiv (2\pi\sigma l)^2+s$.  There are several different possible solutions
depending upon the signs and magnitudes of $\gamma$ and $b$.  
\begin{eqnarray}\label{statsola}
b>0 \quad \gamma>0 && 
r^2_s =  \frac{l^2}{6\gamma}\left[ 1+\sqrt{1-12\frac{q^2\gamma}{\l^2}}\right]
\nonumber \\ 
&& m^2 =  \frac{2}{3}q^2 + 
\frac{l^2}{54\gamma}\left[1+\left(1-12\frac{q^2\gamma}{\l^2}\right)^{3/2}\right] 
\label{statsolb}\\
b>0 \quad \gamma=0 && 
r^2_s =  q  \nonumber \\ && m=q \label{statsolc}\\
b>0 \quad \gamma<0 && 
r^2_s =  \frac{l^2}{6|\gamma|}\left[ -1+\sqrt{1+12\frac{q^2|\gamma|}{\l^2}}\right]
\nonumber \\ && m^2 =  \frac{2}{3}q^2 + 
\frac{l^2}{54|\gamma|}\left[-1+\left(1+12\frac{q^2|\gamma|}{\l^2}\right)^{3/2}\right] 
\label{statsold}\\
b=0 \quad \gamma<0 && 
r^2_s =  \frac{|q|l}{\sqrt{3|\gamma|}} = \frac{2q^2}{3m} \nonumber \\ &&
m^2= \frac {4 q^3\sqrt{|\gamma|}}{3\sqrt{3}l} \label{statsole}\\
b<0 \quad \gamma<0 && 
r^2_s =  \frac{l^2}{6|\gamma|}\left[ 1+\sqrt{1+12\frac{q^2|\gamma|}{\l^2}}\right]
\nonumber \\ && m^2 =  -\frac{2}{3}q^2 + 
\frac{l^2}{54|\gamma|}\left[1+\left(1+12\frac{q^2|\gamma|}{\l^2}\right)^{3/2}\right] 
\label{statsolf}
\end{eqnarray}
For all solutions with $b\neq 0$, $r_s=\frac{3m}{2}\left(b + \sqrt{1-b\frac{8q^2}{9m^2}}
\right)$.

These are the only allowed solutions. Since every solution of the matching condition
must obey $3m=r_s(b+\frac{q^2}{r^2_s})$,
for $b \leq 0$ and $m>0$ there is no solution with zero charge, and
so it will not be possible to pair-create neutral black holes with
non-trivial topology.  However for $b<0$ and $m<0$ static solutions of zero charge
exist with $r_s = 3|m|$.  This will correspond to the pair creation of 
neutral negative mass black holes \cite{negmas}.  Static charged solutions
of negative mass also exist, with $r_s = 3|m| \frac{ 1 + \sqrt{1+e^2} }{2}$.

Non-static solutions of (\ref{match}) describe the creation of 
accelerating black holes and are periodic in the Euclidean time $\tau$.
This period is
\begin{equation}\label{period}
\beta_W = \oint_{r_{\mbox{min}}}^{r_{\mbox{max}}} d\tau 
=  \oint_{r_{\mbox{min}}}^{r_{\mbox{max}}} \frac{dr}{\sqrt{V(V-(2\pi\sigma r)^2)}}
\end{equation}
and is the amount of Euclidean time needed for the wall to 
interpolate between the turning points $r_{\mbox{min}}$ and
$r_{\mbox{max}}$ of the motion (\ref{match}), both of which must be
real and positive.  The wall will intersect itself
unless it moves between the turning points an integral number of times within
the period $\beta_H$. Hence 
\begin{equation}\label{permatch}
\beta_W = \frac{\beta_H}{n}     
\end{equation}
where $n$ is a positive integer.

An analysis of the turning points of the motion involves a consideration of
the function
\begin{equation}\label{turnV}
N - (2\pi\sigma r)^2 \equiv U(r) = -\frac{\gamma}{l^2}r^2 +b -2\frac{m}{r}+\frac{q^2}{r^2}
\end{equation}
for regions $r_{\mbox{min}} \leq r \leq r_{\mbox{max}}$ such that $U(r) \geq 0$. 
This means that $U(r)$ must have at least three roots for positive $r$. 
As with the metric function $V(r)$ in (\ref{rnpot}), there are only two roots
of $U(r)$ if $\gamma < 0$, whereas if $\gamma\geq 0$ there are three roots only
if $b>0$.  An alternative way of seeing this is by realizing that $U(r) \geq 0$ 
implies
\begin{equation}\label{turncond}
{q^2}-2{m}{r} \geq  \frac{\gamma}{l^2}r^4 -b r^2  \qquad .
\end{equation}
The left hand side of this equation is a straight line of slope $-2m$ and intercept
$q^2$, whereas the right hand side is a quartic intersecting the origin and symmetric
about $r=0$.  For negative $\gamma$, the quartic can
intersect the line for positive $r$ at two points at most; 
however in between these regions the inequality is violated unless the points
coincide, in which case inequality is saturated.
For positive $\gamma$ this is also true unless $b$ is positive, in which
case there exists a local maximum of the quartic at the origin, yielding a
region $r>0$ in which the inequality is satisfied (or alternatively saturated at two 
distinct points), provided $\gamma/l^2$ is sufficiently small.  This translates into
a bound on the mass which is $\frac{1}{3\sqrt{3}} \leq \frac{ml}{\sqrt{\gamma}} \leq \frac{\sqrt{2}}{3\sqrt{3}}$.
However from (\ref{bhor}), there are no horizons for $ q > m$. This shifts the
upper bound on $m$ downward, so that $\frac{1}{3\sqrt{3}} \leq \frac{ml}{\sqrt{\gamma}}
\leq \frac{1}{4}$.

Hence the only accelerating black holes that can be pair-created by domain walls 
are those of spherical topology, in either de Sitter or anti de Sitter space, and the
latter scenario is possible only if $\sigma$ is sufficiently large.  Only
static topological black holes can be pair created using domain walls.  This conclusion
holds regardless of the sign of $m$.  Indeed, in order to pair-create negative mass
holes, both $b$ and $\gamma$ must be negative.  

The preceding conclusions are altered if the sign of $q^2$ is reversed, as is
sometimes done in considering electrically charged black holes instead of magnetic
ones \cite{moss}.  Analytically continuing $q\to iq$ will reverse the scenario
described above.  The only accelerating black holes which could be pair-created
would be those of negative mass and genus $g\geq 2$ topology, and would have 
$b<0$ and $\gamma < 0$.  This would violate electromagnetic duality.
However instead of continuing the charge to imaginary values, it is possible to
consider electrically charged instantons in which the electromagnetic field is
pure imaginary on the Riemannian section. This restores duality \cite{cosern,hawross}, 
and leads to the same pair creation scenarios and production rates that the 
magnetically charged holes have.

\section{Topological Black Hole Pair Production}
\label{pair}

Since the condition (\ref{period}) is  only applicable to  black holes of spherical
topology, a more detailed analysis of its validity is identical to that considered
for supergravity domain walls \cite{Chamsugra} with equal negative values of the 
cosmological constant on either side of the wall.  There is a countably infinite
set of instantons which satisfy (\ref{period}) which can mediate the creation of
accelerating spherical black hole pairs from the initial domain wall state.  

All other black hole pairs will be created in a static configuration satisfying one
of the conditions (\ref{statsola}) -- (\ref{statsolf}).  
I shall consider only the pair creation of magnetically charged static
black holes; the electric case yields the same results, but entails 
the incorporation  of an additional surface term  that vanishes in 
the magnetic case \cite{cosern,hawross}.

The Euclidean action for these instantons is
\begin{equation}\label{act1}
I = \int d^4x \sqrt{g} \left(-\frac{R}{16\pi}  
+ \frac{F^2}{16\pi} + {\cal L}_c +{\cal L}_d
\right)
\end{equation}
where ${\cal L}_c$ is the cosmological Lagrangian and ${\cal L}_d$ the
domain-wall Lagrangian. The former may be taken to be that
of the squared field strength of a 3-form or simply the
constant $3\frac{s}{8\pi l^2}$. The domain wall Lagrangian can be
that of a membrane current coupling to the 3-form
\cite{Aur} or that of a scalar field $\Phi$ whose potential 
${\cal V}(\Phi)$ is
everywhere positive \cite{dompair} (and so its Euclidean action is
always negative). There are no boundary terms because the instantons
considered here are compact and without boundary.

Regardless of the mechanism, the Einstein field equations 
applied to (\ref{act1}) yield
\begin{eqnarray}
I &=& \int_{M_e} d^4x \sqrt{g} \left(-\frac{R}{16\pi} + \frac{3s}{8\pi l^2}  
+ \frac{F^2}{16\pi} \right) -\frac{\sigma}{2}\int_W \sqrt{h} d^3x \nonumber\\
 &=& \int d^4x \sqrt{g} \left(-\frac{3s}{8\pi l^2}  
+ \frac{F^2}{16\pi} \right) -\frac{\sigma}{2}\int_W \sqrt{h} d^3x
\label{act2}
\end{eqnarray}
for the Euclidean action, where $\sigma$ is the energy density of the
wall, and the Euclidean section $M_e$ includes the volumes on both sides of
the wall. For a pair of genus $g$ black holes of mass $m$ and charge $q$
\begin{equation}\label{act3}
I(m,q,g) = \beta_H (|g-1| + \delta_{g,1})
\left(-\frac{s(r_s^3-r_H^3)}{l^2}+\frac{q^2}{r_s r_H}(r_s-r_H)
-2\pi\sigma r_s^2\sqrt{N(r_s)}\right)
\end{equation}
where $r_H$ is the location of the outer horizon of the black hole,
$\beta_H$  is given by (\ref{betaH}), and $r_s$ is given by
the relevant equation in (\ref{statsola}) -- (\ref{statsolf}).

The amplitude for pair creation will be approximately $e^{-I/2}$ 
because half the Euclidean section provides an instanton for the pair
creation of black holes.  Hence the rate of pair creation will be 
proportional to the probability $e^{-I}$.  What is physically meaningful 
is a comparison of the creation rate of the combined black hole-wall configuration
relative to the creation rate an appropriate background configuration.
For genus $0$ black holes this background can be taken to be that
of a domain wall with empty adS space on either side, obtained by
gluing two hyperbolic 4-balls along their boundary 3-spheres.
For higher genus black holes this is somewhat problematic, as there
are several choices of a comparative background, depending upon how
one chooses to view the topological black hole creation process.

Within a given topological sector of genus $g$, the domain wall must
have the same topology in order to satisfy the matching conditions.
However a domain wall of this topology cannot be matched to an empty
adS space. The natural background within a given topological sector
would seem to be the $m=q=0$ configuration.  In this case the
relative pair-creation rate would be
\begin{eqnarray}\label{rateg}
\Gamma_g &=& e^{-I(m,q,g)+I(0,0,g)}  \\
       &=& 
\exp\left[|g-1|\left(\frac{2(2\pi\sigma l)^2 -1}{[(2\pi\sigma l)^2-1]^{3/2}}
-\beta_H(\frac{(r_s^3-r_H^3)}{2\l^2}+\frac{q^2}{r_s r_H}(r_s-r_H)
-2\pi\sigma r_s^2\sqrt{N(r_s)})\right)\right] \nonumber
\end{eqnarray}
for $g\geq 2$. For $g=0$ the $m=q=0$ adS space has no event horizon.
The matching condition implies that $2\pi\sigma l = 1$ with an
arbitrary matching radius $r_c$.  The reference creation rate then depends
upon this additional arbitrary parameter, and it is unclear that
a meaningful comparison can be made.

Another alternative would be to compare the creation rate of black holes
of genus $g$ to the creation rate of a pure domain wall with completely
empty genus $g=0$ adS space on either side.  This action $I_B$ for this latter
situation is given by \cite{Chamsugra}
\begin{equation}\label{rateg2}
I_B = \frac{\pi l^2}{3\sqrt{3}}\left(\frac{\sigma l}{2}\sinh^3(\sqrt{3}\frac{r_c}{l})
-\frac{1}{36}\sinh(3\sqrt{3}\frac{r_c}{l})
+ \frac{r_c}{4\sqrt{3}l}\cosh(3\sqrt{3}\frac{r_c}{l})
-\frac{3}{4} e^{-\sqrt{3}r_c/l}\right)
\end{equation}
where $r_c = l/\sqrt{(2\pi\sigma l)^2-1}$.  The pair creation rate is then given by
\begin{equation}\label{rateg3}
\Gamma_{adS} = e^{-I(m,q,g)+I_B}
\end{equation}
for a black hole of genus $g$.

If the cosmological constant is created from
the squared field strength of a 3-form, the creation rate 
is relative to that for creation of a domain wall with no black holes 
or relativisitic 3-form.  In this case the relative rate is given by
\begin{equation}\label{prob}
\Gamma_{3-form} = \exp\left[ (|g-1|+\delta_{g,1})
\left(2\pi\sigma r_s^2\sqrt{V(r_s)}\beta_H -\frac{q^2}{r_s r_+}(r_s-r_H)\beta_H
+ \frac{s(r_s^3-r_H^3)\beta_H}{\l^2}\right)  - \frac{1}{8\pi\sigma^2}\right]
\end{equation}
for the pair creation of  black holes of arbitrary genus.

The above expressions all include the case $s=1$, which correspond to the
production of static black holes in the Reissner-Nordstrom de Sitter case.
If the $\sigma$-dependent terms are omitted, and 
$r_s$ is taken to be the location of the cosmological horizon, 
then the results of  ref. \cite{cosern} are recovered.

\section{Production of Constant Curvature Black Holes}
\label{conscurv}

For $b<0$, the class of metrics (\ref{cas3ban}) may be written as
\begin{equation}\label{ccblack}
ds^2 = (\rho^2-\rho^2_+)\left( -(-\frac{\rho^2_+}{l^2}R^2-k) dT^2 + 
\frac{dR^2}{-\frac{\rho^2_+}{l^2}R^2-k}\right)
+ \frac{d\rho^2}{\frac{\rho^2-\rho^2_+}{l^2}} + \rho^2 d\phi^2
\end{equation}
where without loss of generality I have set $b=-\frac{\rho^2_+}{l^2}$.  The $(R,T)$ sector
is a $(1+1)$ dimensional de Sitter spacetime, with either $\partial/\partial R$ or 
$\partial/\partial T$ being timelike, depending on the sign of $k$ and the
magnitude of $R$. Choosing (again, without loss of generality) $k=-1$ yields
\begin{equation}\label{ccblack2}
ds^2 = l^2(\frac{\rho^2-\rho^2_+}{\rho^2_+})\left( -\sin^2(\theta)dt^2 + 
d\theta^2\right)
+ \frac{d\rho^2}{\frac{\rho^2-\rho^2_+}{l^2}} + \rho^2 d\phi^2
\end{equation}
as an alternate form for (\ref{ccblack}) once the coordinate transformations
$R=\frac{l}{\rho_+}\cos(\theta)$ and $T=\frac{l}{\rho_+}t$ have been carried out.

Banados has recently pointed out \cite{Bancon} that these metrics
(or alternatively the metrics (\ref{ccblack2})) can be understood as 
$(3+1)$ dimensional versions of the $(2+1)$ dimensional BTZ black hole;
in other words, these spacetimes are $(3+1)$ dimensional anti de Sitter
space with identifications differing from those discussed in section
\ref{tads}, but with the property that the identifications produced a
chronological singularity that is hidden behind an event horizon. They
are therefore black holes of constant curvature.

This can be understood in the following way. Consider the standard formulation
of anti de Sitter spacetime, which is that of a hyperboloid
\begin{equation}\label{adshyp}
-x_0^2+x_1^2+x_2^2+x_3^2-x_4^2 = -l^2
\end{equation}
in a flat $(3+2)$ dimensional spacetime described by the  metric
\begin{equation}\label{adsflat}
ds^2 = -dx_0^2+dx_1^2+dx_2^2+dx_3^2-dx_4^2
\end{equation}
which has 4 rotation and 6 boost Killing vectors.  Points along the orbit
of the boost Killing vector $\xi = \frac{\rho_+}{l}(x_i\frac{\partial}{\partial x_4}
+ x_4\frac{\partial}{\partial x_i})$ are identified under a discrete subgroup of
the de Sitter group, where $x_i$ is any one of 
the spacelike coordinates. Points in the region where $\xi^2 \leq 0$ contain
closed timelike curves; hence $\xi^2=0$ is a chronological singularity. This
singularity is a hyperboloid
\begin{equation}\label{adshypsing}
x_0^2-x_1^2-x_2^2 = l^2
\end{equation}
if $x_i=x_3$ has been chosen for the boost direction. The surface for which
\begin{equation}\label{adshyphor}
x_0^2-x_1^2-x_2^2 = 0
\end{equation}
may be regarded as a horizon: within this region timelike geodesics inevitably encounter
the chronological singularity, whereas outside it they do not. The topology of
this surface is a null conoid instead of a null line.

Making the coordinate transformation
\begin{eqnarray}
x_\alpha &=& \frac{2l y_\alpha}{1-y^2}\nonumber\\
x_3 &=& l\frac{1+y^2}{1-y^2}\sinh(\frac{\rho_+}{l}\phi) \label{cctfm}\\
x_4 &=& l\frac{1+y^2}{1-y^2}\cosh(\frac{\rho_+}{l}\phi) \nonumber
\end{eqnarray}
with $\alpha = \{0,1,2\}$ transforms the metric (\ref{adsflat}) to
\begin{equation}\label{cckrus}
ds^2 = \frac{4l^2 dy\cdot dy}{1-y^2} + \rho^2_+\frac{1+y^2}{1-y^2}d\phi^2
\end{equation}
where $dy\cdot dy \equiv dy_1^2+dy_2^2-dy_0^2$ and $y^2\equiv y_1^2+y_2^2-y_0^2$.
The ranges of the coordinates are $-\infty < y_\alpha < \infty$ and
$-\infty < \phi < \infty$, where $ |y^2| < 1$.  The black hole spacetime
results upon identifying $\phi \sim \phi + 2\pi n$. The above metric
can be viewed as the Kruskal form of the black hole (\ref{ccblack2}), 
with the singularity at $y^2=-1$, the horizon at $y^2=0$ and (timelike) 
infinity at $y^2=1$. By setting $\rho=\rho^2_+\frac{1+y^2}{1-y^2}$, and
choosing coordinates so that
\begin{equation}\label{cbltrm}
y_0 = f(\rho)\sin\theta \sinh t \quad y_1 = f(\rho)\sin\theta \cosh t
\quad y_2 = f(\rho)\cos\theta
\end{equation}
where $f(\rho)=\sqrt{(\rho-\rho_+)/(\rho+\rho_+)}$, the metric (\ref{cckrus})
may be shown to be equivalent to the metric (\ref{ccblack2}).

The event horizon is therefore the direct product of a circle with a null
conoid, in contrast to the case I metrics, for which the horizon is
the product of a genus-$g$ 2-surface with a null line.  Analytically continuing
$t\to i\tau$ in (\ref{ccblack2}), the null conoid becomes a 2-sphere, and
the coordinate $\tau$ must be periodic with
period $\beta_\rho = 2\pi l/\rho_+$ to remove conical singularities in the
$(\tau,\rho,\theta)$ section.  The instanton again flares out like a solid bell, but 
with topology $R^3\times S^1$.  Matching two copies of this
manifold together at a radius $r$ at the open end of their bells
yields a Riemannian section with topology  $S^3\times S^1$. There is a
single domain wall of topology $S^2\times S^1$ and two bolts
of topology $S^1$ where the Killing field $\frac{\partial}{\partial \tau}$ vanishes.
The nucleation surface $\Sigma$ which joins the Lorentzian and Riemannian sections
is located along the $\tau=0$ and $\tau=\beta_\rho/2$ segments.

The matching condition differs from that given for the case I metrics.  The 
Lanczos conditions yield
\begin{equation}\label{ccmatch}
\left. \left[ h^\prime(\rho) + \frac{h(\rho)}{\rho}\right]\right|_{\rho=\rho_s} 
= 4\pi\sigma  
\end{equation}
where $h^2(\rho) = \frac{\rho^2-\rho^2_+}{l^2}$ and only static solutions
at $\rho=\rho_s$ are being considered.  Solving (\ref{ccmatch}) for $\rho_s$ yields
\begin{equation}\label{rhomatch}
\rho^2_s = \frac{\rho^2_+}{2}\left[1 + \frac{2\pi\sigma l}{\sqrt{(2\pi\sigma l)^2-1}}\right]
\end{equation}
where $\rho_s$ must be positive. Matching is only possible provided $l\sigma$ is
sufficiently large, $2\pi l\sigma > 1$. 

These instantons are uncharged, and so the action is easily calculated to be
\begin{eqnarray}\label{ccact}
I_{cc} &=& \int d^4x \sqrt{g} \frac{3}{8\pi l^2}  
       -\frac{\sigma}{2}\int_W \sqrt{h} d^3x \nonumber \\
  &=& \frac{l\beta_\rho}{2\rho^2_+}\sqrt{(\rho^2_s -\rho^2_+)^3}    
       - \rho_s\frac{\pi\sigma l^2\beta_\rho}{\rho^2_+}(\rho^2_s -\rho^2_+)    
\end{eqnarray}
with $\rho_s$ given by (\ref{rhomatch}).  The production rate of these constant
curvature black holes is then given by 
\begin{equation}\label{ratecc}
\Gamma_{adS} = e^{-I_{cc}+I_B}
\end{equation}
relative to the creation rate of a pure domain wall with completely
empty adS space on either side, with $I_B$ given by (\ref{rateg2}).

\section{Conclusions}\label{conc}

By requiring the cosmological c-metric to be free of conical singularities in the
$(x,\phi)$ section, a large variety of candidate black hole spacetimes emerged. There
were three classes of such spacetimes. The first (case I) included black holes
whose event horizons are compact 2-surfaces with topology of arbitrary genus $g$. For
$g=0$, these black holes are the usual Reissner-Nordstrom (anti) de Sitter type.
For $g\geq 1$, they are all asymptotically Reissner-Nordstrom anti de Sitter
($\Lambda < 0$), with the entire spacetime inheiriting the topology of the event horizon.  
For genus $g\geq 2$ solutions with both positive and negative mass are permitted.
The second class (case II) all have $\Lambda < 0$, and the event horizons are non-compact
2-surfaces.  The third class (case III) of metrics are all asymptotic to the
constant curvature black holes (\ref{ccblack}), but contain naked singularities
unless $m=q=0$. For $m=q=0$ these metrics are the constant curvature black
holes discussed by Banados \cite{Bancon}.

Black hole pairs of both the case I and case III classes may be produced using
domain walls, although this is not necessary if $\Lambda > 0$ \cite{cosern}. For
case I metrics, if
$\Lambda < 0$, then the only allowed solutions are static unless the topology
of the event horizon is spherical ($g=0$).  The production rates of these 
topological black holes calculated in section \ref{pair} are for black holes of a given
genus $g$.  In general, the larger the genus, the more suppressed the production
rate. More generally one could compute the rate for producing black hole
pairs of all possible topologies.  This involves a simple sum over the genus
which yields
\begin{equation}
\Gamma = e^{-I(m,q,0)} + e^{-I(m,q,1)} + \frac{e^{-2\hat{I}(m,q)}}{1-e^{-\hat{I}(m,q)}}
\end{equation}
for the (unnormalized) inclusive production rate, where 
\begin{equation}
\hat{I}(m,q) = 2\pi\sigma r_s^2\sqrt{V(r_s)}\beta_H -\frac{q^2}{r_s r_+}(r_s-r_H)\beta_H
- \frac{(r_s^3-r_H^3)\beta_H}{\l^2} \quad .
\end{equation}

In $N=1$ supergravity theories, domain walls naturally arise as boundaries between
regions of isolated vacua of the supergravity matter fields. There is no {\it a-priori}
reason to exclude walls of a given topology.  A wall of a specified topology will
in general be quantum mechanically unstable to pair creation of black holes of the
same toplogy, as the preceding arguments in this paper demonstrate.  The extension 
of these arguments to situations in which rotation, dilatonic couplings, and
charged domain walls are included  remain interesting open questions.

\section*{Acknowledgements}

This work was supported by the Natural Sciences and Engineering Research 
Council of Canada. Many of the calculations in this paper
were carried out using the GrTensor package with MAPLE V.4 software.


\begin{thebibliography}{99}
\bibitem{dgkt}H.F. Dowker, J.P. Gauntlett, D.A. Kastor and J. Traschen,
 Phys. Rev.  D {\bf 49}, 2909  (1994).
\bibitem{cosern}R.B. Mann and S.F. Ross, Phys. Rev. {\bf D52}, 2254
(1995).
\bibitem{strpair}S.W. Hawking and Simon F. Ross, Phys. Rev. 
Lett. {\bf 75} (1995) 3382;  R. Emparan 
Phys. Rev. Lett. {\bf 75} (1995) 3386 ; D. Eardley, G. Horowitz,
D. Kaster and J. Traschen, Phys. Rev. Lett. {\bf 75} (1995) 3390. 
\bibitem{dompair}R.R. Caldwell, A. Chamblin and G.W. Gibbons,
``Pair Creation of Black Holes by Domain Walls'', hep-th/9602216.
\bibitem{ernst}F. J. Ernst, J. Math. Phys. {\bf 17}, 515 (1976).
\bibitem{dggh}H.F. Dowker, J.P. Gauntlett, S.B. Giddings and G.T. Horowitz,
 Phys. Rev. D {\bf 50}, 2662 (1994).
\bibitem{entar}S.W. Hawking, G.T. Horowitz and S.F. Ross, Phys. Rev. {\bf D 51} 4302
(1995), gr-qc/9409013.
\bibitem{cqgtop}R.B. Mann, Class. Quant. Grav. {\bf 14} L109 (1997),
gr-qc/9607071.
\bibitem{gen} J.F. Plebanski and M. Demianski, Ann. Phys. {\bf 98}, 98
(1976).
\bibitem{btz}M. Banados, C. Teitelboim and J. Zanelli,
Phys. Rev. Lett. {\bf 69}, (1992) 1849; M. Banados, M. Henneaux, C. Teitelboim and J. Zanelli,
 Phys. Rev. {\bf D} {\bf 48}, (1993) 1506.
\bibitem{wendy}W.L. Smith and R.B. Mann, ``Formation of Topological Black Holes
from Gravitational Collapse'', WATPHYS-TH97/02, gr-qc/9703007.
\bibitem{BLP}D. Brill, J. Louko and P. Peldan, ``Thermodynamics of
$(3+1)$-dimensional Black Holes with Toroidal or Higher Genus Horizons'',
USITP 97-6, gr-qc/9705012.
\bibitem{vanzo}L. Vanzo, ``Black Holes with Unusual Topology'', UTF-400,
gr-qc/9705004.
\bibitem{amin}S. Aminneborg, I Bengtsson, S. Holst and P. Peldan,
Class. Quant. Grav. {\bf 13}, 2707 (1996),
gr-qc/9604022.
\bibitem{negmas}R.B. Mann, ``Black Holes of Negative Mass'', WATPHYS-TH97/03,
gr-qc/9705007.
\bibitem{Bancon}M. Ba\~nados, ``Constant Curvature Black Holes'', gr-qc/9703040.
\bibitem{Chamsugra}A. Chamblin and J.M.A. Ashbourn-Chamblin, ``Black Hole
Pairs and Supergravity Domain Walls'', NSF-ITP-96-150, hep-th/9612014.
\bibitem{Cmetric}W. Kinnersley and M. Walker, Phys. Rev. D {\bf  2}, 
1359 (1970).
\bibitem{dan}J.D. Christensen and R.B. Mann, Class. Quant. Grav {\bf 9} 
(1992) 1769.
\bibitem{bplane}R.G. Cai and Y.Z. Zhang, Phys. Rev. {\bf D54} 4891 (1996), 
gr-qc/9609065.
\bibitem{bjm}J.D. Brown, J. Creighton and R.B. Mann, Phys. Rev. D {\bf 50}
6394 (1994). 
\bibitem{His}W.A. Hiscock  Phys. Rev. {\bf D35}, (1987) 1161.
\bibitem{Aur}A. Aurelia, R. Kissack, R.B. Mann and M. Spallucci,
  Phys. Rev. {\bf D35} (1987) 2961.
\bibitem{moss} F. Mellor and I. Moss, Phys. Lett. {\bf B222}, 361 (1989); {\it 
ibid}, Class. Quant. Grav. {\bf 8}, 1379 (1989).
\bibitem{hawross} S.W. Hawking and S.F. Ross, Phys. Rev. {\bf D52} 5865 (1995), 
hep-th/9504019.



\end{thebibliography}
\end{document}